\documentclass{article}
\usepackage{ltwol2e}

\def\be{\begin{equation}}
\def\ee{\end{equation}}
\def\bea{\begin{eqnarray}}
\def\eea{\end{eqnarray}}

\usepackage{epsfig}			

\begin{document} 

\onecolumn
\pagestyle{empty}
\large
\rightline{SLAC-PUB-8278}
\rightline{19-Nov-1999}
\begin{center}
\vspace*{2.8cm}
{\LARGE\bf Review of \mbox{\boldmath $B\bar B$} Mixing
\renewcommand{\thefootnote}{\fnsymbol{footnote}}Results\footnote{Work 
supported by U.S. Department of Energy contract DE-AC03-76SF00515.}}\\
\vspace*{6.0ex}
{\Large Mourad Daoudi}\\
\vspace*{1.5ex}
{\it Stanford Linear Accelerator Center\\
Stanford, CA 94309}\\
\vspace*{1.2in}
{\Large\bf Abstract}\\
\end{center}
\vspace{0.2cm}
A review of $B\bar B$ mixing results at the end of July 
1999 is presented. Emphasis is put on recent measurements of 
$\Delta m_d$ and $\Delta m_s$. For $\Delta m_d$, the new world average 
is $\Delta m_d = 0.473\pm 0.016\ \rm ps^{-1}$. For $\Delta m_s$, the new 
world average 95\% CL limit is $12.4\ \rm ps^{-1}$, with a sensitivity of 
$14.2\ \rm ps^{-1}$. Other related results are covered very
briefly.\ \\
\noindent

\vspace*{2.2in}
\begin{center}
{\large
Invited talk at 8th International Symposium on Heavy Flavour Physics\\
Southampton, UK, July 25--29, 1999}
\end{center}
\normalsize
\newpage
\pagestyle{plain}



\title{Review of {$B\bar B$} Mixing Results}

\author{Mourad Daoudi}
\address{Stanford Linear Accelerator Center, MS-78, P.O.Box 4349, 
        Stanford, CA 94309, USA\\
	E-mail: daoudi@SLAC.Stanford.EDU}


\twocolumn[\maketitle\abstracts{ 
A review of $B\bar B$ mixing results at the end of July 
1999 is presented. Emphasis is put on recent measurements of 
$\Delta m_d$ and $\Delta m_s$. For $\Delta m_d$, the new world average 
is $\Delta m_d = 0.473\pm 0.016\ \rm ps^{-1}$. For $\Delta m_s$, the new 
world average 95\% CL limit is $12.4\ \rm ps^{-1}$, with a sensitivity of 
$14.2\ \rm ps^{-1}$. Other related results are covered very
briefly.}]
\vspace{2.2in}


\section{Introduction}
The main motivation for performing $B\bar B$ mixing measurements lies in the
determination of the $CKM$ matrix element $V_{td}$, which represents one of
the constraints on the Unitarity Triangle.
\par
\begin{figure}[h]
\begin{center} 
{\epsfig{file=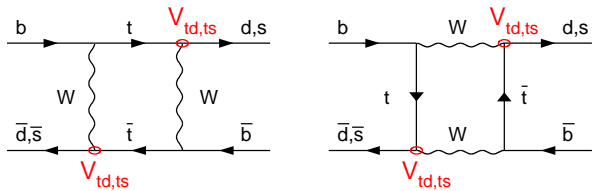,width=3.2in}%
        \caption{$B\bar B$ mixing diagrams.}%
	\label{boxd}}
\end{center}
\end{figure}

$V_{td}$ is accessible experimentally through the box diagrams of 
figure \ref{boxd}, by
measuring the mass difference $\Delta m_d$ in $B_d$ mixing.
$\Delta m_d$ and $V_{td}$ are related by:
\begin{eqnarray*}
\Delta m_d & = & {{G_F^2}\over{6\pi^2}}\ m_b\ m_t^2\ F\left(
{m_t^2\over m_W^2}\right)\ \eta_{QCD} \\
& &
\times\ \ B_{B_d}\ f_{B_d}^2\ \left|V_{tb}^*\ V_{td}\right|^2.
\end{eqnarray*}
Similarly $B_s$ mixing provides a measurement of $V_{ts}$. 
\par
While $\Delta m_d$
is measured with very good precision, the determination of $V_{td}$ is limited
by theoretical uncertainties in the decay constant $f_{B_d}$ and the bag 
factor $B_{B_d}$. However, in the ratio $\Delta m_s$/$\Delta m_d$ most 
hadronic uncertainties cancel\ \cite{flynn}:
\begin{eqnarray*}
{\Delta m_s\over\Delta m_d} & = &
{{\eta_{B_s}M_{B_s}\ f_{B_s}^2 B_{B_s}} \over
{\eta_{B_d}M_{B_d}\ f_{B_d}^2 B_{B_d}} }\ 
\left| {V_{ts}\over V_{td}} \right|^2 \\
& = & (1.14\pm 0.08)^2\ \left| {V_{ts}\over V_{td}} \right|^2.
\end{eqnarray*} 
With $|V_{ts}|\simeq |V_{cb}|$, a precise measurement of both 
$\Delta m_d$ and $\Delta m_s$ provides a strong constraint on $V_{td}$.
This underlines the importance of $B_s$ mixing measurements.
\begin{figure}[h]
\begin{center}
{\epsfig{file=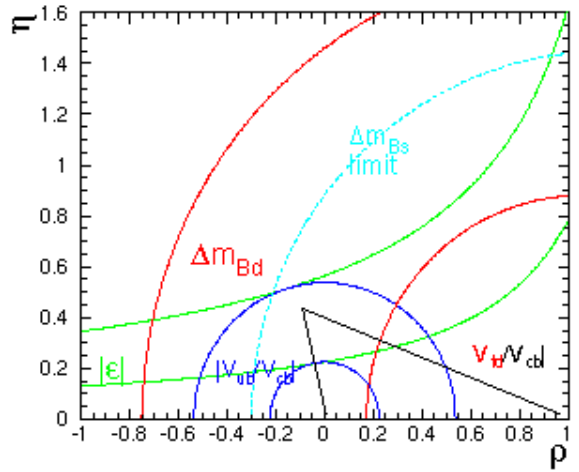,width=3.2in}%
\caption{Illustration of the various constraints on the Unitarity Triangle.}%
	\label{unitri}}
\end{center}
\end{figure}
\par
In the Wolfenstein parametrization the CKM matrix is defined by four
parameters, two of which are quite well known, 
$\lambda = \sin\theta_C = 0.2205$ and
$A \simeq |V_{cb}|/\lambda^2 \sim 0.8$, and two that are not well 
determined, $\rho$ and $\eta$ whose
values define the apex of the Unitarity Triangle.
In the $\rho\eta$-plane, $V_{td}$ is represented by a circle centered at (1,0).
It provides one of three constrains on the Unitarity Triangle, the other two
being $V_{ub}$ and $\epsilon_K$ (figure \ref{unitri}).
$\Delta m_d$ and $\Delta m_s$ are given by
\begin{eqnarray*}
\Delta m_d & \propto & A^2\lambda^6\left[(1-\rho)^2 + \eta^2\right], \\
\Delta m_s & \propto & A^2\lambda^4,
\end{eqnarray*}
in terms of $\rho$, $\eta$, $\lambda$, and $A$. 
Note that, unlike $\Delta m_d$, $\Delta m_s$ has no dependence on $\rho$ and
$\eta$. Therefore, a precise measurement of $\Delta m_s$ provides in essence
a measure of the product $f_{B_s}\sqrt{B_{B_s}}$. 
\par
Another important aspect is the fact that $B_s$ mixing is complementary to the
$CP$ violation related measurements that will be performed at $B$ factories 
in the near future. The measurements of $V_{td}$ and $\sin 2\beta$ being 
essentially orthogonal, they will provide together an excellent constraint on
the apex of the Unitarity Triangle in the $B$ system alone. Whereas, 
measurements of $V_{ub}$ and $\sin 2\beta$ alone will not be sufficient to
provide as good a constraint.
\par
Improvements in lattice calculations continue to be made. For the latest
developments in the determination of the decay constant $f_{B_d}$ and the bag 
factor $B_{B_d}$, see the talk by S. Hashimoto at this 
conference\ \cite{hashimoto}. Further details can be found in 
Refs.\ \cite{flynn,ali}. It is generally accepted that the product 
$f_{B_d}\sqrt{B_{B_d}}$ lie\ \cite{babar} in the range $[160,240]$ MeV.
\subsection{Time-integrated mixing measurements}
The probability that a $B^0$ meson mix into a $\bar{B^0}$ or remain unmixed
as a function of proper time is:
$$
{\cal P}_{B^0\rightarrow \bar{B^0},B^0}(t) = \Gamma e^{-\Gamma t}\
(1\mp\cos\Delta m t) / 2.
$$
The time-integrated mixed fraction is given by:
$$
\chi = {(\Delta m/\Gamma)^2\over{2[1+(\Delta m/\Gamma)^2]}}.
$$
At the $\Upsilon(4S)$ where $B_d$ mesons are produced essentially at rest,
$\Delta m_d$ can extracted from a measurement of $\chi_d$. Because
$\Delta m_s$ is large ($1/\lambda^2\approx 20$ times larger than $\Delta m_d$),
it can be obtained only in a time-dependent mixing measurement 
($\chi_s\sim 0.5$), which makes it rather very challenging.
\par
At the $Z^0$ resonance, where all $B$ species are produced, a measurement of
$\bar\chi = f_d\chi_d + f_s\chi_s$ allows the extraction of 
$f_s\equiv f(B_s)$, the fraction
of $b$ quarks that hadronize into a $B_s$.

\subsection{Time-dependent mixing measurements}
The requirements for a time-dependent mixing measurement are: 1) a precise 
reconstruction of the proper time of the $B$ meson, 2) a determination of the
flavor $b$ or $\bar b$ of the $B$ meson both at production (initial-state tag)
and at decay (final-state tag).
\par 
A variety of initial-state tags are used, they fall in three categories:
1) vertex and jet charge, high-momentum lepton and kaon, in the opposite
hemisphere, 2) fragmentation kaon or pion in the same hemisphere, 3) polarized
forward-backward asymmetry. The latter is unique to SLD, it exploits the 
parity-violating $Zb\bar b$ coupling and the presence of a highly polarized
electron beam at the SLC.
\par 
The most widely used final-state tag is the sign of a high-$P_t$ lepton. Other
tags include, for example, the sign of a partially or fully reconstructed 
$D_s^\pm$ meson, and the dipole charge (see below). 

\section{$B$ productions fractions}
The LEP B Oscillations Working Group\ \cite{lepbosc} has compiled a new value 
for $\bar\chi = 0.1186\pm 0.0048 $. Furthermore, with the inclusion of recent
$\Delta m_d$ measurements by CLEO and CDF, mentioned below, a new world
average value for $\chi_d$ was derived: $\chi_d = 0.176\pm 0.009$.
\par
A measurement of the charged $B$ branching ratio made by 
DELPHI\ \cite{delphibr}
recently resulted in an increase of the $B_u$ production fraction: 
$f(B_u)\equiv f(B_d) = (40.5\pm 1.2)\%$. While, the average $b$-baryon
production fraction has decreased slightly: $(9.5\pm 2.0)\%$.
As a result, the new $B_s$ production fraction becomes:
$$ f(B_s) = (9.6\pm 1.3)\%. $$
This value is in good agreement with a separate determination using the
measurement of the branching ratio ${\cal B}(b\rightarrow B_s)\times
{\cal B}(B_s\rightarrow D_s^-l^+\nu X)$ which yields: 
$ f(B_s) = (9.7\pm 2.3)\%. $
The new value of $f(B_s)$ is somewhat smaller compared to that evaluated
a year ago at the ICHEP'98 conference: $(10.8\pm 1.3)\%$\ \cite{lepbosc}. This 
will have the effect of reducing one's sensitivity to $\Delta m_s$, compared 
to one's previous estimate assuming a larger value of $f(B_s)$.

\section{$B_d$ mixing measurements}
Over the last year there have been two new measurements of $\Delta m_d$: one
from CLEO and the other from CDF.
\par
The CLEO time-integrated measurement\ \cite{cleo} is based
on a 4.2 Million $B\bar B$ event sample. The final-state $b$ flavor is tagged
using the sign of a $slow$ pion in the partially-reconstructed decay 
chain: $\bar B_d^0\rightarrow D^{*+}\pi^- (\rho^-)$, 
$D^{*+}\rightarrow D^0\pi_s^+$. The initial-state is tagged with a 
high-momentum lepton in the rest of the event. 
A value of $\chi_d = 0.195\pm 0.026\pm 0.032$
is obtained, from which the following result for $\Delta m_d$ is extracted:
$$ \Delta m_d = 0.512\pm 0.053(stat)\pm 0.032(syst)\ \rm ps^{-1}. $$
This result is competitive with those from LEP, CDF,
and SLD obtained in time-dependent analyses.
\par
At CDF, a new time-dependent measurement was performed\ \cite{cdf}. It consists
of the reconstruction of a $\bar B_d$ decaying to either a $D^+$ or a 
$D^{*+}$:
\begin{eqnarray*}
 \bar B_d^0 &\rightarrow& D^+X, \ \  D^+\rightarrow K^-\pi^+\pi^+,\\
 \bar B_d^0 &\rightarrow& D^{*+}X, \ D^{*+}\rightarrow D^0\pi^+,
      \ D^0\rightarrow K^-\pi^+.
\end{eqnarray*}
\begin{figure}[h]
\begin{center}
{\epsfig{file=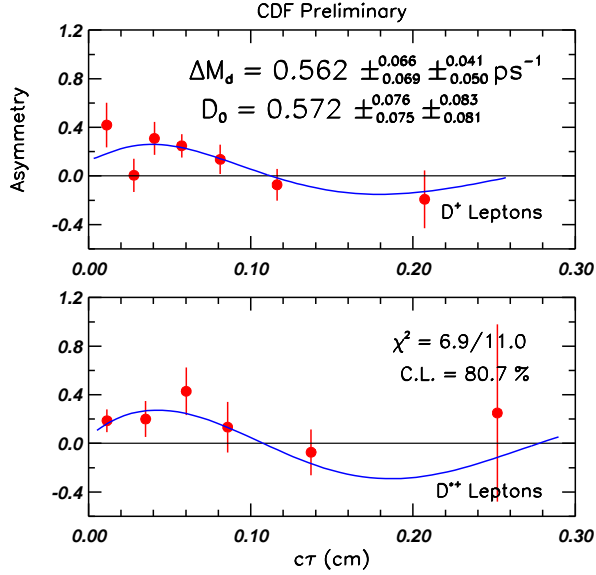,width=3.2in}%
\caption{Mixing asymmetry vs. proper time distribution for the $D^+$ and 
         $D^{*+}$ samples at CDF.}%
	\label{cdfdmd}}
\end{center}
\end{figure}
The sign of the $D^{(*)\pm}$ tags the $b$ flavor of the $B_d$ at decay while 
its flavor at production is given by a high momentum lepton in the opposite 
hemisphere.
\begin{figure}[t]
{\hbox{\hskip -2cm\epsfig{file=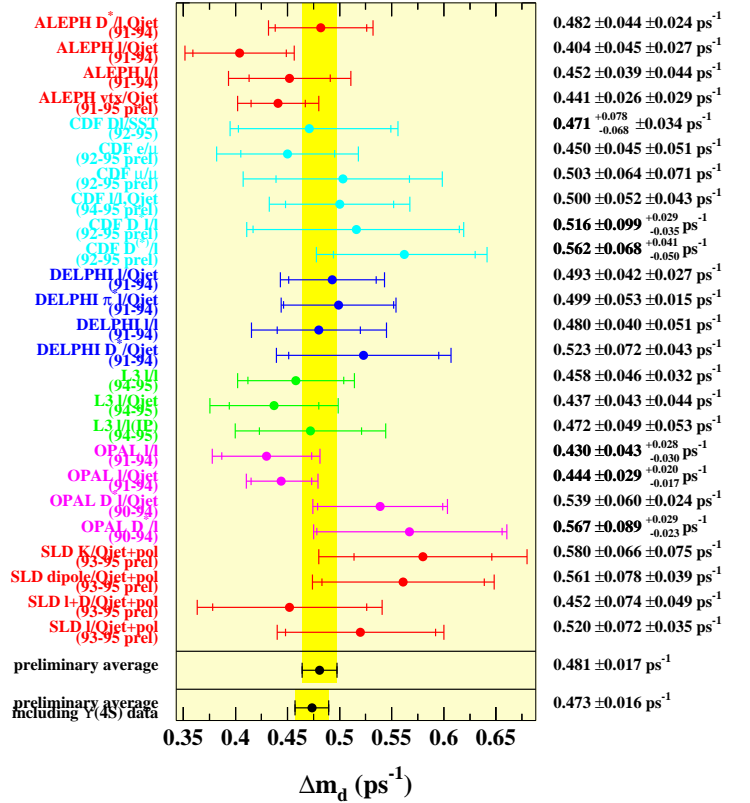,width=4.7in}}%
\caption{Measurements of $\Delta m_d$.}%
	\label{dmd}}
\end{figure}
A nice feature of this analysis is the fact that $\Delta m_d$ is extracted
from a maximum likelihood fit in which the tagging dilution factor ($D_0$)
is also fitted. Effectively, the mistag rate is determined directly from
the data.
figure \ref{cdfdmd} shows the mixing asymmetry distribution
$A(t) = (N_{mix}-N_{unmix}) / N_{tot}$ as a function of proper time, for
the $D^+$ and $D^{*+}$ samples separately. The combined fit results in:
$$
\Delta m_d = 0.562\pm 0.068(stat)\ ^{+0.041}_{-0.050}(syst)\ \rm ps^{-1}.
$$
The value obtained from the fit for the dilution factor is: 
$D_0 = 0.572\pm 0.080(stat)^{+0.083}_{-0.081}(syst)$.
\par
figure \ref{dmd} contains a compilation of about 25 measurements of 
$\Delta m_d$ from the four LEP experiments, CLEO, CDF, and SLD, all very 
consistent with one another. The new world average value is\ \cite{lepbosc}: 
$$ \Delta m_d = 0.473\pm 0.016\ \rm ps^{-1}. $$
Further improvements in $B_d$ mixing are expected from CLEO and SLD which
have relatively large portions of data not yet analyzed.
\section{$B_s$ mixing measurements}
Over the last few years, a lot of efforts have been spent on $B_s$ mixing.
Before reviewing some of the recent analyses, let us mention the first 
experimental limits on the ratio
$\Delta\Gamma_s/\Gamma_s$ which provide indirect upper limits
on $\Delta m_s$. Unlike $B_d$, recent calculations indicate that for $B_s$ 
the ratio $\Delta\Gamma/\Gamma$ is sizeable\ \cite{beneke}.  
Experimentally, the limits are derived by fitting in a $B_s$-enriched sample
the proper time distribution to a sum of two exponentials. The best limit 
comes from DELPHI\ \cite{delphibs}:
$$ \Delta\Gamma_s/\Gamma_s < 0.42\ \rm at\ 95\%\ CL.$$
Limits on $ \Delta\Gamma_s/\Gamma_s $ were also set by L3 and CDF.
Though the above limit does not provide a strong constraint on $\Delta m_s$ 
($< 110$ ps$^{-1}$), future improvements are expected as additional and 
perhaps improved measurements are made.
\par
Three ingredients are required for a $\Delta m_s$ measurement, besides a 
large data sample. They are a high $B_s$ enrichment ($f_s$), a small mistag 
rate ($\eta$), and a good proper time resolution ($\sigma_t$). This is 
illustrated by the following expression which gives the statistical 
significance to a $\Delta m_s$ signal\ \cite{moser}:
$$ S = \sqrt{N/2}\ f_s\ (1-2\eta)\ {\rm exp}\left[
-(\Delta m_s\ \sigma_t)^2/2\right]. $$ 
$N$ is the total number of $B$ decays in the sample.
The proper time resolution can be expressed as:
$\sigma_t^2 = \left(\sigma_l/\beta\gamma c\right)^2 
            + t^2\ \left(\sigma_{\beta\gamma}/\beta\gamma\right)^2$, where
$\sigma_l$ is the decay length resolution and $\sigma_{\beta\gamma}$ is the
boost resolution. Clearly, for a large value of $\Delta m_s$, all the
sensitivity is at very short proper time, and the decay length resolution is
the dominant term there.
\par
A widely used approach to reconstruct a $B_s$ meson and tag its final-state
$b$ flavor is to intersect a high-$P_t$ lepton with a downstream vertex which
may be a fully- or partially-reconstructed $D_s^\pm$ meson. 
Sometimes the $B_s$ fraction in the sample is enhanced by requiring that a 
kaon or a $\phi$ meson be present in this vertex.
\par
The extraction of a $\Delta m_s$ limit relies on the so-called 
{\sl Amplitude Fit} method\ \cite{moser}, where the $(1-\cos\Delta m_st)$ 
term in the above expression for the time-dependent mixed probability is
replaced by $(1-A\cos\Delta m_st)$. Values of the amplitude parameter $A$ 
are scanned for each trial value of $\Delta m_s$ in a maximum likelihood fit.
At the true mixing frequency, the amplitude $A = 1$, and is $=0$
for all other values of $\Delta m_s$. The 95\% CL limit corresponds to the
smallest value of $\Delta m_s$ for which the measured amplitude and its error 
satisfy the relation: $A + 1.645\ \sigma_A = 1$. Similarly, 
the {\sl sensitivity} of a given measurement corresponds to the smallest 
value of $\Delta m_s$ for which $1.645\ \sigma_A = 1$ holds. 
\par
A total of eleven analyses have been performed by ALEPH, CDF, DELPHI, OPAL,
and SLD. 
This summer, OPAL and DELPHI have produced updates on their $B_s$ mixing 
results. These are described below. For completeness, important results from
other experiments are also reviewed.

\subsection{ALEPH inclusive lepton analysis}
This is the most sensitive $B_s$ mixing measurement to date\ \cite{alephbs}. 
$B_s$ mesons
are reconstructed as the intersection of a high-$P_t$ lepton and an inclusive
$D$ vertex. The final-state flavor is given by the lepton sign, while the
initial-state tag relies on vertex charge or a high-momentum lepton in the
\begin{figure}[h]
\begin{center}
{\epsfig{file=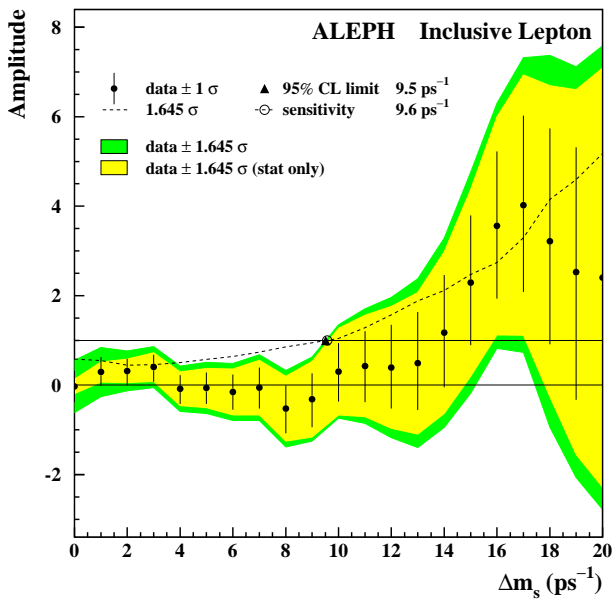,width=3.2in}%
\caption{Amplitude distribution for ALEPH's inclusive lepton analysis.}%
	\label{alephdl}}
\end{center}
\end{figure}
opposite hemisphere, or a fragmentation kaon in the decay hemisphere. A total
of 33,023 decays are selected. The decay length resolution is modeled by a
double-Gaussian distribution with a core (82\%) $\sigma_l = 280\ \mu$m and
a tail (18\%) $\sigma_l=1060\ \mu$m. The relative boost resolution is 
estimated by $\sigma_{\beta\gamma}/\beta\gamma = 7.1\%$ in the core (72\%) 
and $\sigma_{\beta\gamma}/\beta\gamma = 21\%$ in the tails (28\%). 
figure \ref{alephdl} shows the amplitude vs. $\Delta m_s$ distribution obtained
in this analysis. The derived limit is: $\Delta m_s>9.5$ ps$^{-1}$ at 95\% CL,
with a sensitivity of $9.6$ ps$^{-1}$.
Combined with two other analyses ($D_sl$\ \cite{alephdsl} and 
$D_sh$\ \cite{alephdsh}), 
ALEPH's sensitivity becomes $10.6\ \rm ps^{-1}$ and the overall limit is:
$$ \Delta m_s > 9.6\ \rm ps^{-1}\ at\ 95\%\ CL. $$
\subsection{OPAL inclusive lepton update}
OPAL uses a rather sophisticated technique to reconstruct in 3-D a $B_s$ 
vertex that contains a high-$P_t$ lepton\ \cite{opal}. The final-state is
tagged with the lepton sign, while the initial-state is given by either jet 
charge or a high-$P_t$ lepton in the opposite hemisphere. In this update, OPAL
\begin{figure}[h]
\begin{center}
{\epsfig{file=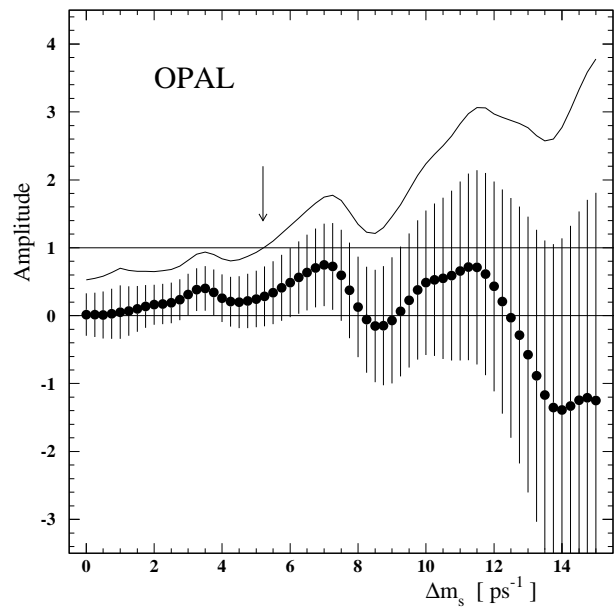,width=3.2in}%
\caption{Amplitude distribution for OPAL's inclusive lepton analysis.}%
	\label{opalil}}
\end{center}
\end{figure}
included their 1995 data and made substantial improvements to their jet charge
technique. The final analysis is based on 47,109 single-lepton and 6,031
di-lepton events. The amplitude vs. $\Delta m_s$ distribution is shown in
figure \ref{opalil}.
The sensitivity is $7.0\ \rm ps^{-1}$ and the limit is:
$$ \Delta m_s > 5.2\ \rm ps^{-1}\ at\ 95\%\ CL. $$
\subsection{DELPHI exclusive $B_s$ and $D_s^\pm h^\mp$ update}
DELPHI has made the first attempt at an exclusive $B_s$ 
reconstruction\ \cite{delphibs}.
This approach has the advantage of an excellent proper time resolution, 
with the $B_s$ boost being determined exactly.
The following decay channels are utilized:
\begin{eqnarray*}
B_s & \rightarrow & D_s^-\pi^+\ (a_1^+), \\
B_s & \rightarrow & {\bar D^0}K^-\pi^+\ (a_1^+),
\end{eqnarray*}
where the $D_s^-$ and $\bar D^0$ are reconstructed in 6 and 2 decay modes, 
respectively.
From the full data sample corresponding to 3.6 Million hadronic $Z$'s,
11 fully- and 33 partially-reconstructed $B_s$ decays are selected with
a corresponding $B_s$ purity of 70\% and 45\%, respectively. 
With such a small efficiency this analysis
by itself has no sensitivity to $\Delta m_s$. However, because of its 
excellent proper time resolution it provides a non-negligible contribution 
at large $\Delta m_s$ values. 
\par
DELPHI choose to combine it with the more inclusive 
$D_s^\pm h^\mp$ ($D_s\rightarrow\phi\pi, K^*K$) analysis\ \cite{delphibs} 
where 2953 candidates are selected with a $B_s$ purity of 40\%.
The overall sensitivity so obtained is $3.2\ \rm ps^{-1}$ and the 95\% CL
limit is $4.0\ \rm ps^{-1}$.
\subsection{DELPHI $D_s^\pm l^\mp$ update}
This is overall the second most-sensitive $B_s$ mixing 
analysis\ \cite{delphibs}. A total of 436 decays with a fully-reconstructed 
$D_s$ are selected in the decay modes
$D_s^+\rightarrow\phi\pi^+, K^*K,^+, \phi l^+\nu$. Another
\begin{figure}[h]
\begin{center}
{\epsfig{file=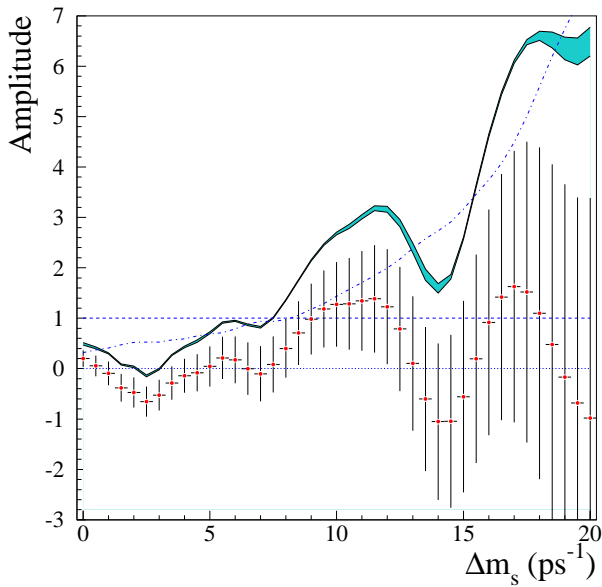,width=3.2in}%
\caption{Amplitude distribution for DELPHI's $D_sl$ analysis.}%
	\label{delphidsl}}
\end{center}
\end{figure}
441 candidates with a partially-reconstructed (missing $\gamma/\pi^0$)
$D_s^+\rightarrow\phi l^+X$ are also selected. The $B_s$ content in the
two sub-samples is estimated to be $230\pm 18$ and $41\pm 12$ $B_s$'s,
respectively. Clearly, the $D_s$ sign provides the final-state tag, while
a complex initial-state package including 9 discriminant variables gives
the initial-state $b$ flavor with a 78\% purity. The decay length resolution
in this analysis is parametrized with a core (86\%) Gaussian  with 
$\sigma_l=220\ \mu$m and a tail (14\%) Gaussian with $\sigma_l=870\ \mu$m.
Whereas, the boost resolution is characterized by a $5.4\%$ relative residual 
in the core (78\%) and $17.2\%$ in the tails (22\%). 
The 95\% CL limit extracted from this analysis is $7.4\ \rm ps^{-1}$ with a
sensitivity of $8.2\ \rm ps^{-1}$. This is illustrated by the amplitude vs. 
$\Delta m_s$ distribution shown in figure \ref{delphidsl}.
The combined limit from all the DELPHI analyses is:
$$ \Delta m_s > 5.0\ \rm ps^{-1}\ at\ 95\%\ CL $$
with an overall sensitivity of $9.7\ \rm ps^{-1}$.
\subsection{CDF $\phi l\ /\ l$ analysis}
Using their di-lepton trigger data, CDF\ \cite{cdfbs} perform a partial 
reconstruction of the $B_s$ in the decay chain 
$B_s\rightarrow D_sl\nu X\nu\rightarrow\phi l\nu X$.
A sample of 1068 candidates is selected with a $B_s$ purity of 61\%. A second
high-momentum lepton in the opposite hemisphere tags the initial-state flavor.
\begin{figure}[h]
\begin{center}
{\epsfig{file=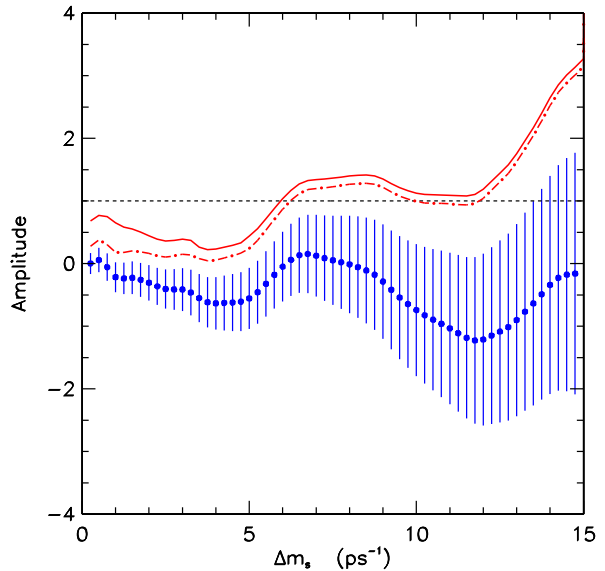,width=3.2in}%
\caption{Amplitude distribution for CDF's $\phi l/l$ analysis.}%
	\label{cdfdms}}
\end{center}
\end{figure}
This analysis results in a limit of:
$$ \Delta m_s > 5.8\ \rm ps^{-1}\ at\ 95\%\ CL, $$
and a sensitivity of $5.1\ \rm ps^{-1}$, as shown by the amplitude distribution
of figure \ref{cdfdms}.
\subsection{SLD $B_s$ mixing results}
Using a data sample equivalent to $350k$ hadronic $Z^0$'s, SLD has performed
three $B_s$ mixing analyses, referred to as {\sl Lepton+D},
{\sl Lepton+Tracks}, and {\sl Dipole}\ \cite{thom}. 
The number of selected candidates
in the three analyses is 2352, 8864, and 8211, respectively. In the first two,
the final-state tag is provided by the high-$P_t$ lepton, while in the last
one it is given by the dipole charge, defined as the charge difference 
between the tertiary and secondary vertices multiplied by the distance 
separating the two vertices.
\begin{figure}[h]
\begin{center}
{\epsfig{file=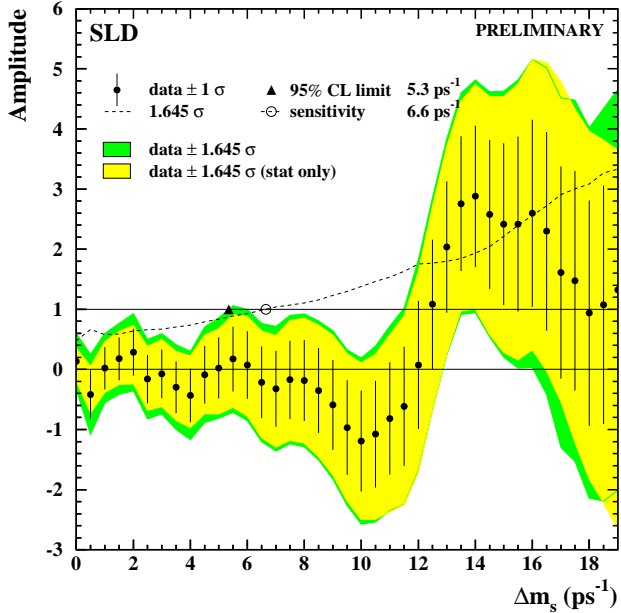,width=3.2in}%
\caption{Combined amplitude distribution for the Lepton+D, Lepton+Tracks, 
and Dipole analyses at SLD.}%
	\label{sld}}
\end{center}
\end{figure}
The initial-state tag is given primarily by the polarized forward backward
asymmetry, and by jet charge, vertex charge, a high-momentum lepton or kaon
in the opposite hemisphere. An excellent decay length resolution is achieved
at SLD, characterized by a core (60\%) Gaussian with 
$\sigma_l = 105 - 130\ \mu$m and a tail (40\%) Gaussian with 
$\sigma_l = 330 - 550\ \mu$m. Good boost resolution is also realized:
$\sigma_{\beta\gamma}/\beta\gamma = 7 - 9\ \%$ for the core (60\%) and 
$\sigma_{\beta\gamma}/\beta\gamma = 22 - 30\ \%$ for the tails (40\%).
The combined amplitude distribution for the three analyses is shown in
figure \ref{sld}. An overall sensitivity of $6.6\ \rm ps^{-1}$ is obtained,
and two separate intervals in $\Delta m_s$ are excluded at the 95\% CL:
\begin{eqnarray*}
 & \Delta m_s & < 5.3\ \ \rm and\\
 6.0 < & \Delta m_s & < 11.5\ \rm ps^{-1}.
\end{eqnarray*}
\subsection{Summary of $\Delta m_s$ results}
\begin{figure}[h]
\begin{center}
{\epsfig{file=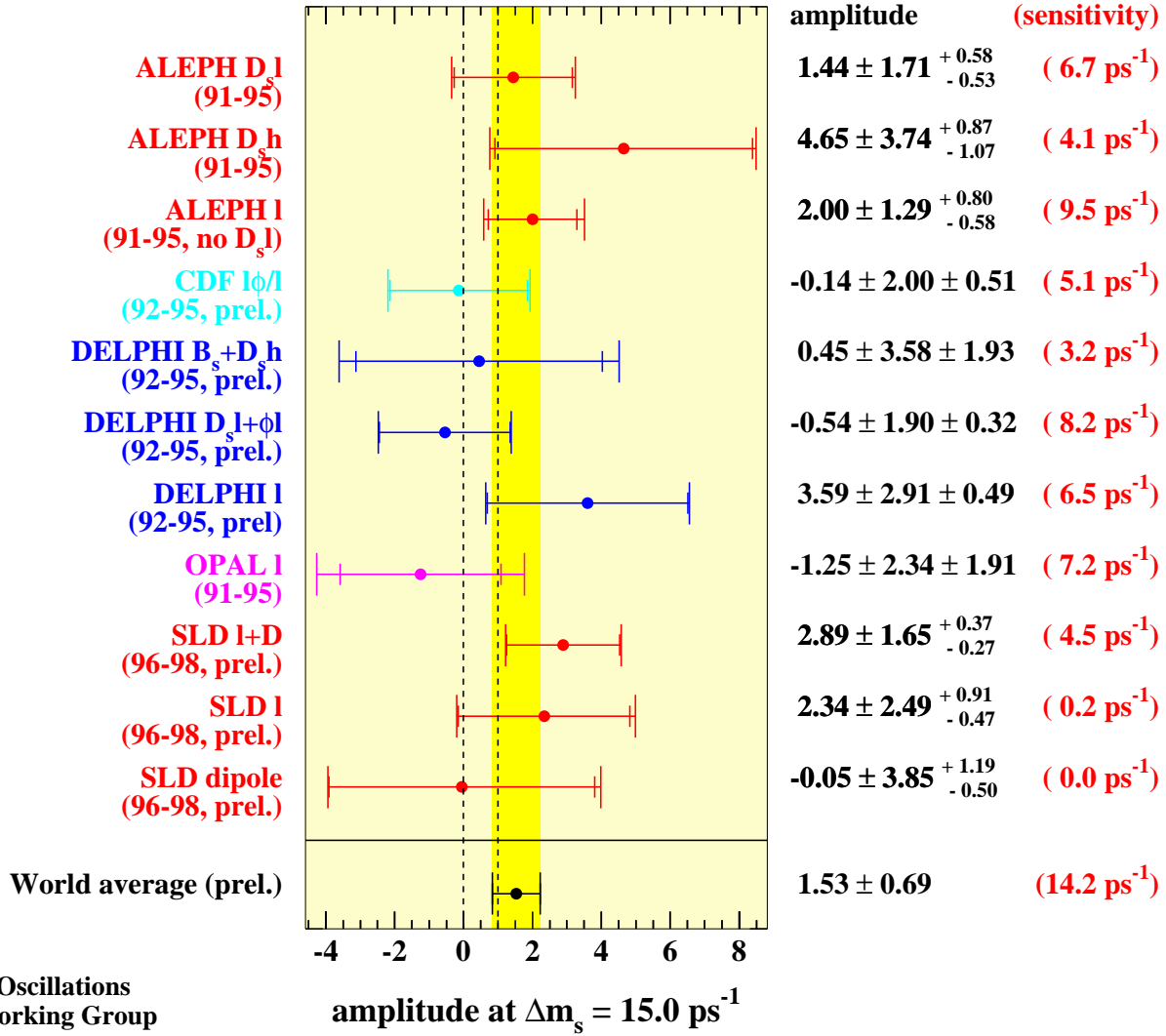,width=6.5in}%
\caption{Measured amplitude at $\Delta m_s=15\ \rm ps^{-1}$ for each 
         analysis.}%
	\label{amp15}}
\end{center}
\end{figure}

In the following, a summary as of the end of July 1999 of all $B_s$ mixing 
results compiled by the LEP B Oscillations Working Group\ \cite{lepbosc} is 
given. figure \ref{amp15} gives the measured amplitude and its error at a 
$\Delta m_s$ value of 15 ps$^{-1}$, as well as the sensitivity of each 
analysis. The world average amplitude at $\Delta m_s = 15\ \rm ps^{-1}$ is
$2.2\ \sigma$ away from zero ($1.53\pm 0.69$).
\par
\begin{figure}[t]
\begin{center}
{\epsfig{file=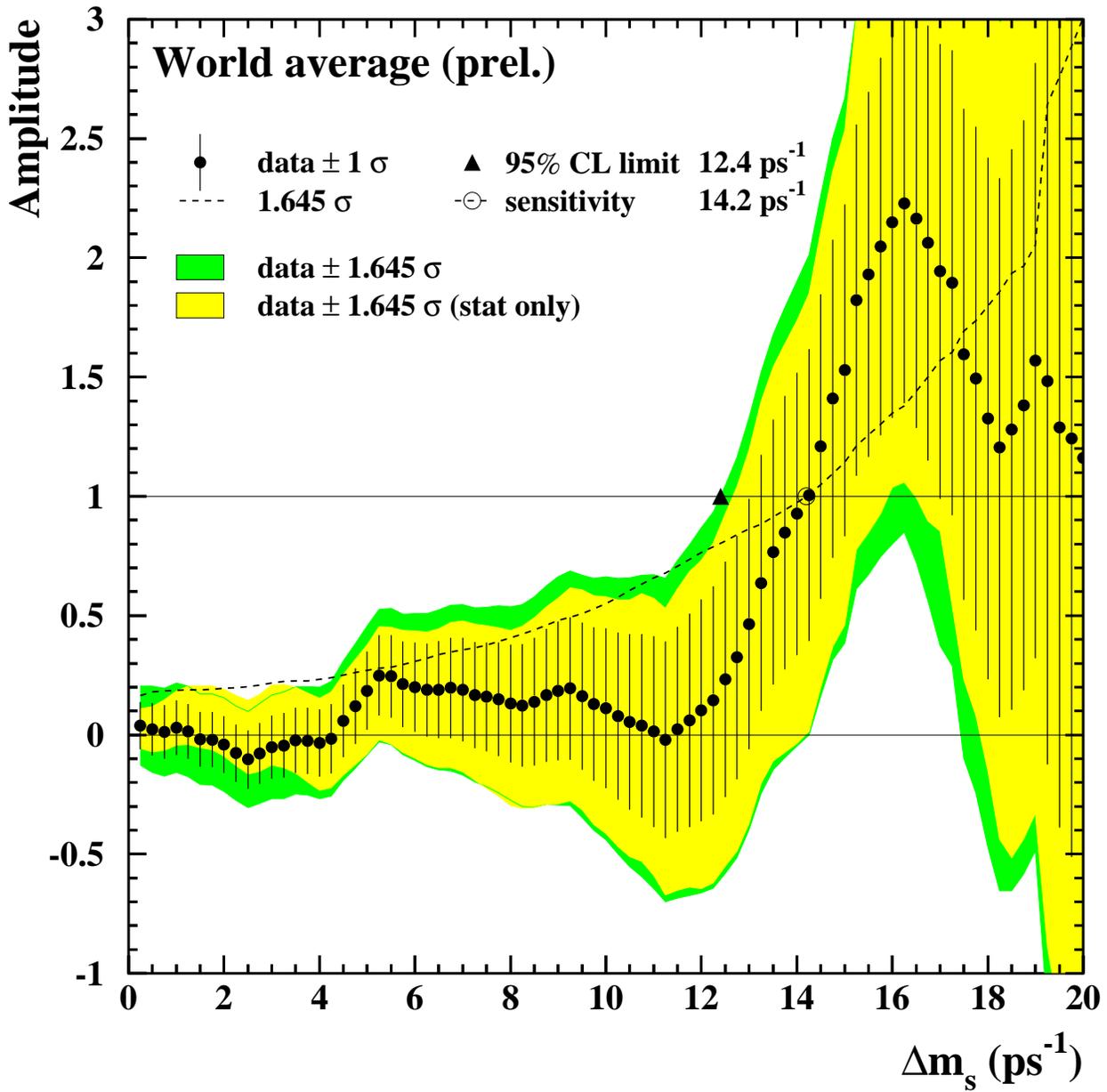,width=6.5in}%
\caption{World average amplitude vs. $\Delta m_s$ distribution.}%
	\label{ampavg}}
\end{center}
\end{figure}

The world average amplitude vs. $\Delta m_s$ distribution is shown in
figure \ref{ampavg}. The overall sensitivity is $14.2\ \rm ps^{-1}$ and the
combined limit:
$$ \Delta m_s > 12.4\ \rm ps^{-1}\ at\ 95\%\ CL. $$
Note that this limit does not take into account the new smaller value of the
$B_s$ production fraction, as discussed above. Compared to last year's
results at the ICHEP'98 conference\ \cite{parodi}, the $\Delta m_s$ limit is 
unchanged, while
the overall sensitivity has improved slightly (up from $13.8\ \rm ps^{-1}$).
\begin{figure}[h]
\begin{center}
{\epsfig{file=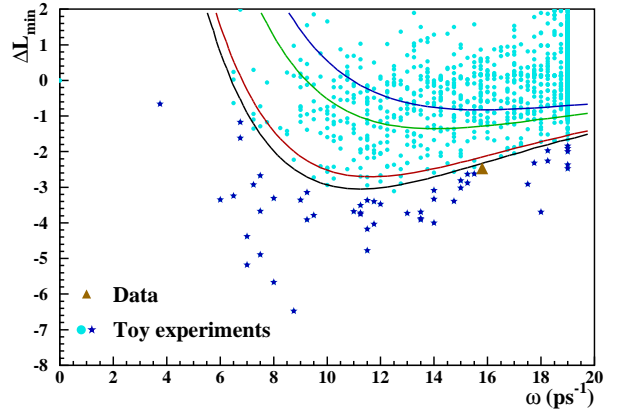,width=3.2in}%
\caption{Likelihood profiles for 2000 toy Monte Carlo samples with 
        $\Delta m_s=150\ \rm ps^{-1}$. The triangle represents the experimental
        data as of winter'99.}%
	\label{abbaneo}}
\end{center}
\end{figure}
The contribution of the limit on $\Delta m_s$ to the determination of CKM
matrix is summarized in Ref.\ \cite{plaszczynski}.
\par
The {\sl bump} seen in the amplitude distribution at around $15\ \rm ps^{-1}$
has been the subject of a lot interest. In a recent contribution, D.~Abbaneo
and G.~Boix\ \cite{boix} proposed a procedure to estimate the probability 
that the observed
structure is due to a pure statistical fluctuation. They ran 2000 fast Monte 
Carlo experiments with a large value of $\Delta m_s$ (150 ps$^{-1}$), each 
experiment being equivalent to the world sensitivity as of last winter 
(Moriond'99)\ \cite{boscjune}.
They convert the measured amplitudes into likelihood profiles as a function
of $\Delta m_s$. From these profiles which are shown in figure \ref{abbaneo}, 
they estimate the probability that the bump in the data (represented by the 
triangle in the figure) to originate from a fluctuation in a no-signal sample
to be $\sim 3 - 5\ \%$.
\clearpage
\ \vskip 1.0in
\clearpage
\clearpage
\subsection{SLD status and future prospects}
Final $B_s$ mixing results from SLD are expected in the near future. Since
Moriond'99, substantial improvements have been achieved, they come
from the following three sources:\\
\indent $\bullet$ Improved tracking resolution.\\
\indent $\bullet$ Improved charge dipole reconstruction.\\
\indent $\bullet$ Addition of two new analyses.
\par
The design resolution with the pixel vertex detector VXD3 has been realized.
It is characterized by a track impact parameter resolution of $7.8\ \mu$m
in the $r\phi$-plane and $9.7\ \mu$m in the $rz$-plane. Furthermore, the
precise location of the micron-size SLC beam spot is determined with an
uncertainty of $4\ \mu$m.
\par
Improvements in the SLD's topological vertexing algorithm\ \cite{jackson}
have resulted in
a significantly better dipole tag purity (80\% compared to 73\% previously).
The dipole tag, which is unique to SLD, exploits the charge flow in the
cascade decay $b\rightarrow c$, and is made possible due to the ability to
reconstruct (topologically) well-separated $B$ and $D$ vertices.
\begin{figure}[h]
\begin{center}
{\epsfig{file=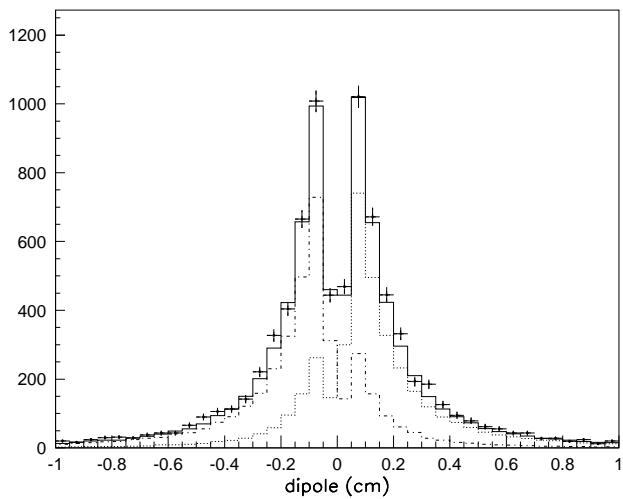,width=3.2in}%
\caption{SLD's dipole charge distribution for data (points) and Monte Carlo
        (histogram). The dotted and dashed histograms represent the $b$- and
        $\bar b$ components, respectively.}%
	\label{dipole}}
\end{center}
\end{figure}
The dipole charge is reconstructed as the charge difference between the $B$
and $D$ vertices multiplied by the distance separating them. It is shown in
figure \ref{dipole}, where the $b$-quark and $\bar b$-quark components are 
represented by the dotted and the dashed histograms, respectively.
\par
One of the two new analyses at SLD, called the $D_s$ {\sl + Tracks} 
analysis,  relies on the exclusive reconstruction of a $D_s^+$ in the 
$\phi\pi^+$ and $K^{*0}K^+$ decay modes. It provides both a high $B_s$ 
purity of 33\%, and an excellent decay length resolution 
$\sigma_l=46\ \mu$m in the core (60\%) and $\sigma_l=158\ \mu$m in the 
tails (40\%). 
\par
In the second new SLD analysis, called the {\sl Lepton + Kaon} analysis, 
final-states containing an opposite sign lepton -- kaon pair are selected. 
It is aimed primarily at isolating the semileptonic decays: 
$B_s\rightarrow D_s^{**-}l^+\nu$, 
$D_s^{**-}\rightarrow K^-{\bar D^{(*)0}}$. Decays of $B^\pm$ and $B_d$ mesons
produce lepton -- kaon pairs with the same sign and are therefore suppressed.
Thus, resulting in an enhancement of the $B_s$ purity. This analysis is also
sensitive to the decays $B_s\rightarrow D_s^-l^+\nu$, $D_s^-\rightarrow\phi X$,
$\phi\rightarrow K^-K^+$, in particular with the additional requirement of
a second kaon in the $B_s$ vertex.
The $B_s$ purity that is achieved is 26\%, while a very good decay length
resolution is also obtained with $\sigma_l=71\ \mu$m in the core (60\%) and
$\sigma_l=330\ \mu$m in the tails (40\%).
Both these new analyses have the disadvantage of a small overall efficiency.
They have a small sensitivity to $\Delta m_s$, but provide a substantial
contribution at high values of $\Delta m_s$ when combined with the other
SLD analyses.
\begin{figure}[h]
\begin{center}
{\epsfig{file=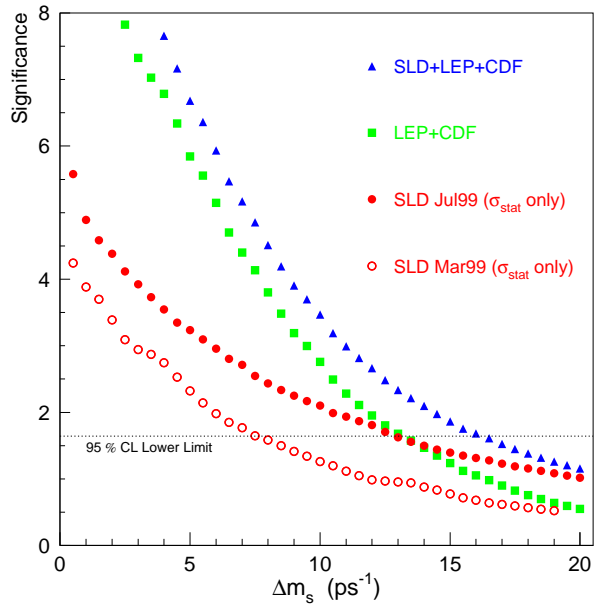,width=3.2in}%
\caption{Estimated improvement in SLD's sensitivity and its impact on the
        world average.}%
	\label{signif}}
\end{center}
\end{figure}
\par
With these improvements, SLD's overall $\Delta m_s$ sensitivity is estimated 
to increase to $\sim 13\ \rm ps^{-1}$, and to dominate the world average for
$\Delta m_s > 13\ \rm ps^{-1}$. The combined LEP+CDF+SLD sensitivity becomes
$\sim 16\ \rm ps^{-1}$. This is illustrated in figure \ref{signif}.
\par
\begin{figure}[h]
\begin{center}
{\epsfig{file=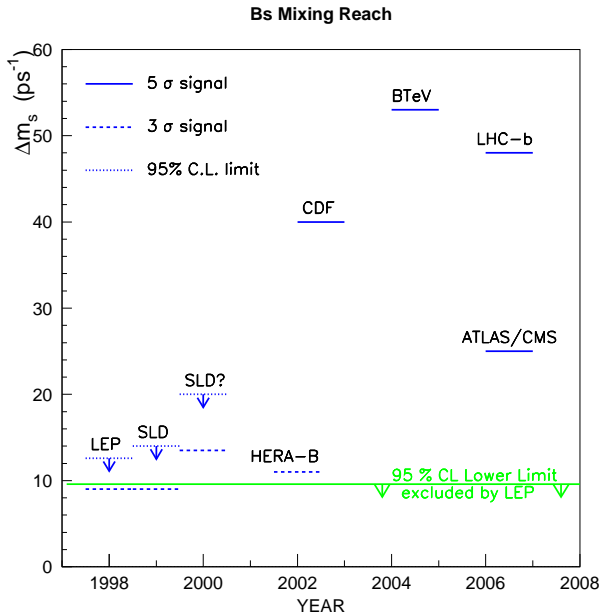,width=3.2in}%
\caption{Expected $B_s$ mixing reach for future-generation $B$ physics
        experiments.}%
	\label{worldscene}}
\end{center}
\end{figure}

Several new experiments\ \cite{gerndt,jesik,harnew} specifically geared for 
$B$ physics, $CP$ violation in particular, are planned for the next decade.
Their expected reach for $B_s$ mixing is shown in figure \ref{worldscene}. It 
is anticipated that CDF will be the first experiment to measure $\Delta m_s$, 
with a sensitivity of $40\ \rm ps^{-1}$. Note that SLD's reach
could have been extended to $\Delta m_s=20\ \rm ps^{-1}$. However, the run
extension that was requested in order to achieve that was not approved due
to lack of funding.
\section{Conclusions}
We have presented a review of $B$ mixing results as of the end of July 1999.
For $B_d$ mixing, with a new measurement from CDF, the new world average 
for $\Delta m_d$ becomes:
$$ \Delta m_d = 0.473\pm 0.016\ \rm ps^{-1}. $$
\par
In $B_s$ mixing, updates from OPAL and DELPHI were produced recently. The new
world average average limit is:
$$ \Delta m_s > 12.4\ \rm ps^{-1}\ at\ 95\%\ CL, $$
and the new world average sensitivity is $14.2\ \rm ps^{-1}$. This, without 
taking into account the smaller value of the $B_s$ production fraction, which 
is evaluated by the LEP B Oscillations Working Group to be:
$$ f(B_s) = (9.6\pm 1.3)\%. $$
 \par
A relatively significant ($\sim 2 \sigma$) {\sl bump} persists in the 
amplitude spectrum at around $\Delta m_s=15\ \rm ps^{-1}$.
An attempt at estimating the probability that this may be due to a statistical
fluctuation was performed by G. Abbaneo and G. Boix (using winter'99 data).
Their result is a probability of $\sim 3 -5\ \%$.
\par
We have also seen the first attempts at constraining $\Delta m_s$ from above,
by setting a lower limit on the ratio $\Delta\Gamma_s/\Gamma_s$. The best limit
comes from DELPHI:
$$ \Delta\Gamma_s/\Gamma_s < 0.42\ \rm at\ 95\%\ CL. $$
\par 
In the near future, final results from SLD are expected, with substantial
improvements coming from a better tracking resolution, an improved dipole
charge reconstruction, and the inclusion of two new analyses. As
of this writing, a partial update from SLD\ \cite{sldaug99} has already 
been presented at the Lepton-Photon'99 conference. In fact the world 
average $\Delta m_s$ limit and sensitivity compiled 
there\ \cite{blaylock,lepbosc} supersede the above quoted values.
\par
In the longer term, many $B$ physics experiments are planned well into the
next decade. With a $\Delta m_s$ reach extending over a very wide range, a
precise measurement of $\Delta m_s$ will be performed by these experiments.

\section*{Acknowledgments}
I would like to thank my colleagues David Jackson, John Jaros, and St\'ephane
Willocq for very fruitful discussions.
\clearpage



\section*{References}
\begin{thebibliography}{999}
\bibitem{flynn}J.M. Flynn, C.T. Sachrajda, [hep-lat/9710057].\\
         Also in Heavy Flavours II, pp. 402-452
         (World Scientific, Singapore, 1997), A.J. Buras, M. Lindner, Eds.
\bibitem{ali}A. Ali, D. London, \it Eur. Phys. J. 
         \rm {\bf C9} (1999) (68);
         A. Ali, D. London, \it Nucl. Phys. \rm{\bf 54A} 
         \it (Proc. Suppl.)\rm (1997) 297.
\bibitem{hashimoto}S. Hashimoto, ``Summary of lattice results for decay 
         constants, mixing, etc.'', in these proceedings.
\bibitem{babar}``The BABAR Physics Book'', P.F. Harrison, H.R. Quinn, Eds.,
         [SLAC-R-504] (1998).
\bibitem{lepbosc}LEP B Oscillations Working Group, see\\
         http://www.cern.ch/LEPBOSC/\\
        (and links/references therein).
\bibitem{delphibr}DELPHI Collaboration, [DELPHI 99-104, CONF 291].
\bibitem{cleo}CLEO Collaboration, see\\
         http://www.lns.cornell.edu/public/CLEO\\
         /analysis/results/B-mixing/.
\bibitem{cdf}CDF Collaboration, see\\
         http://www-cdf.fnal.gov/physics/new\\
         /bottom/cdf4526/cdf4526.html.
\bibitem{moser}H.G. Moser, A. Roussarie, \it Nucl. Inst. Meth.
         \rm{\bf A 384} (1997) 491.
\bibitem{beneke}See for example\\
         M. Beneke et al., \it Phys. Lett. \rm{\bf B459} (1999) 631.
\bibitem{delphibs}DELPHI Collaboration, [DELPHI 99-109, CONF 296].
\bibitem{alephbs}ALEPH Collaboration, \it Eur. Phys. J. 
         \rm{\bf C7} (1999) 553, 
         [hep-ex/9811018].
\bibitem{alephdsl}ALEPH Collaboration, \it Phys. Lett. 
         {\bf B377} (1996) 205.
\bibitem{alephdsh}ALEPH Collaboration, \it Eur. Phys. J.
         \rm{\bf C4} (1998) 367.
\bibitem{opal}OPAL Collaboration, [CERN-EP-99-085], June 1999, 28pp., 
         [hep-ex/9907061].
\bibitem{cdfbs}CDF Collaboration, \it Phys. Rev. Lett.
         \rm{\bf 82} (1999) 3576.
\bibitem{thom}J. Thom, SLD Collaboration, Proceedings of Beauty'99, 
         Bled, Slovenia, June 1999, to appear.
\bibitem{parodi}F. Parodi, Proceedings of the 29th International
         Conference on High Energy Physics, Vancouver 1998, vol. 2, 
         pp. 1148-1154.  
\bibitem{plaszczynski}S. Plaszczynski, ``Overall determination of the CKM
         Matrix'', in these proceedings.\\
         See also:\\
         F. Parodi, P. Roudeau, A. Stocchi, [hep-ex/9903063]; 
         S. Mele, \it Phys. Rev.
         \rm{\bf D59} (1999) 113011. 
\bibitem{boix}G. Boix, D. Abbaneo, \it J. High Energy Phys.
         \rm{\bf 08} (1999) 004, [hep-ex/9909033].
\bibitem{boscjune}LEP B Oscillations Working Group, ``Combined Results on 
         $B^0$ Oscillations: Results from Winter 1999 Conferences'',
         [LEPBOSC 99/1], June 1999.
\bibitem{jackson}D.J. Jackson, \it Nucl. Inst. Meth.
         \rm{\bf A 388} (1997) 247.
\bibitem{gerndt}E. Gerndt, ``Status of HERA-B'', in these proceedings.
\bibitem{jesik}R. Jesik, ``b-physics potential of Tevatron Run II'', 
         in these proceedings.
\bibitem{harnew}N. Harnew, ``Prospects for LHCb, BTeV, ATLAS, CMS'', 
         in these proceedings.
\bibitem{sldaug99}SLD Collaboration, [SLAC-PUB-8225], Aug. 1999, 25pp. 
\bibitem{blaylock}G. Blaylock, Proceedings of Lepton-Photon'99, 
         Stanford, CA, August 1999, to appear.

\end{thebibliography}
\end{document}